\documentclass[useAMS,usenatbib]{mn2e}
\usepackage[dvips]{graphicx}
\usepackage{longtable}

\title[Reflections on Reflexions]{Reflections on Reflexions: 
	II. Effects of Light Echoes on the luminosity and spectra of Type Ia
	Supernovae}
\author[F. Patat et al.]
  {Ferdinando Patat$^1$, Stefano Benetti$^2$, Enrico Cappellaro$^2$ \&  
	Massimo Turatto$^2$\\ \
  $^1$European Southern Observatory, Karl-Schwarzschild
  str. 2, D-85748 Garching bei M\"unchen, Germany.\\ 
  $^2$ INAF - Osservatorio Astronomico di Padova,
  V. dell'Osservatorio, 5, I-35122, Padova, Italy \\ 
   }

\def\dm15{$\Delta m_{15}(B)$}

\begin{document}

\date{Accepted.......;  Received .......}

\pagerange{\pageref{firstpage}--\pageref{lastpage}} \pubyear{2002}

\maketitle

\label{firstpage}

\begin{abstract}
In this paper we present and discuss the effects of scattered light
echoes (LE) on the luminosity and spectral appearance of Type Ia
Supernovae (SNe). After introducing the basic concepts of LE spectral
synthesis, by means of LE models and real observations we investigate
the deviations from pure SN spectra, light and colour curves, the
signatures that witness the presence of a LE and the possible
inferences on the extinction law. The effects on the photometric
parameters and spectral features are also discussed. In particular,
for the case of circumstellar dust, LEs are found to introduce an
apparent relation between the post-maximum decline rate and the
absolute luminosity which is most likely going to affect the well
known Pskowski-Phillips relation.
\end{abstract}

\begin{keywords}
supernovae: general - supernovae: individual (SN 1991T) - supernovae:
individual (SN 1998bu) - supernovae: individual (SN 2001el) - ISM:
dust, extinction nebulae
\end{keywords}

%
%________________________________________________________________

\section{Introduction}
\label{sec:intro}

Due to the crucial role played by Type Ia SNe in modern cosmology (see
for example the review by \citealt{leibundgut}), it is essential to
understand the systematics which might affect the recent conclusions
about the expansion rate of the Universe. The availability of larger
telescopes and the renewed interest for these objects as cosmological
probes has opened the possibility of studying in great detail their
evolution, an investigation which in the past was feasible for the
closest events only. In these last few years, in fact, numerous Ia
have been followed with an extensive time coverage as late as one year
past maximum light, leading to very interesting results (see for
example \citealt{benetti05}).

In two cases, namely those of SN~1991T \citep{schmidt} and SN~1998bu
\citep{cappellaro,jason}, the spectra at advanced phases were clearly
dominated by a phenomenon which had nothing to do with the explosion
mechanism but rather with the neighboring environment.  In both cases
this very atypical behaviour could be explained with the contamination
by SN light scattered into the line of sight by dust present in the
surroundings of the explosion. The presence of such a phenomenon, also
known as light echo (hereafter LE), was later confirmed by high
spatial resolution HST imaging (\citealt{sparks,garnavich}). Other
cases have been discovered for core-collapse SNe like SN~1987A
\citep{crotts89}, SN~1993J
\citep{sugerman02,liu}, SN~1999ev \citep{maund} and SN~2003gd 
\citep{sugerman05,vandyk}. In the most recent years, LEs have been searched 
also in the case of historical SNe \citep{martino, krause, rest}.

With the purpose of studying this effect and its possible applications
to the analysis of the SN environments and their relation with the SN
progenitors, we have started a series of papers on this topic. The
first of them (\citealt{paperI}; hereafter Paper I) was dedicated to
the general characteristics of LEs, with particular attention to the
effects of self absorption on LE luminosity, spectral and colour
appearance. In this and subsequent papers we will instead present and
discuss applications to known and test cases.  In principle, LEs can
affect all types of SNe. Due to the young stellar population from
which they are supposed to be originated, core-collapse SNe are of
course favored with respect to thermonuclear explosions. However, both
due to their high homogeneity (which makes the analysis much simpler)
and their wide use in cosmology, here we will concentrate on Type Ia only.
In particular, the present work deals with the effects of LEs on their
luminosity and spectra, in order both to help the observers in disentangling
intrinsic features from LE contamination and to understand whether LEs
can affect the conclusions reached on their nature.

The paper is organized as follows. After giving a short introduction
to the basic concepts of LE modeling in Section~\ref{sec:basic}, in
Section~\ref{sec:speclib} we describe the spectral library used in our
calculations. The results are then presented and discussed in
Section~\ref{sec:effects} for two different dust geometries, while
Section~\ref{sec:ext} is devoted to the implications on the extinction
law. In Section~\ref{sec:discuss} we discuss our findings which are
briefly summarized in Section~\ref{sec:conclusion}.

\section{Basic concepts}
\label{sec:basic}

The LE phenomenon has been presented and discussed in a number of
publications and we refer to them the interested reader (see for
example the recent review by \citealt{sugerman} and the references
therein).  The detailed description of the formation of a LE spectrum
has been outlined in Paper I and, therefore, here we will only shortly
recap the basic concepts.

In single scattering approximation, the LE spectrum at a given time
$t$ can be expressed as

\begin{equation}
\label{eq:spectrum}
S_{LE}(\lambda,t) = \int_0 ^t S^0(\lambda, t-t^\prime) f(\lambda, t^\prime) 
\; dt^\prime
\end{equation}

where $S^0(\lambda,t)$ is the intrinsic source (SN) spectrum and
$f(\lambda, t)$ is a function that describes the physical and
geometrical properties of the dust. If we define $\tau_d(\lambda)$ as
the optical depth to the SN along the line of sight and $S(\lambda,t)$
is the SN spectrum clean of LE contamination, then

\begin{displaymath}
S^0(\lambda,t)=S(\lambda,t) \; e^{\tau_d(\lambda)} 
\end{displaymath}

\begin{figure}
\centering
\resizebox{\hsize}{!}{\includegraphics{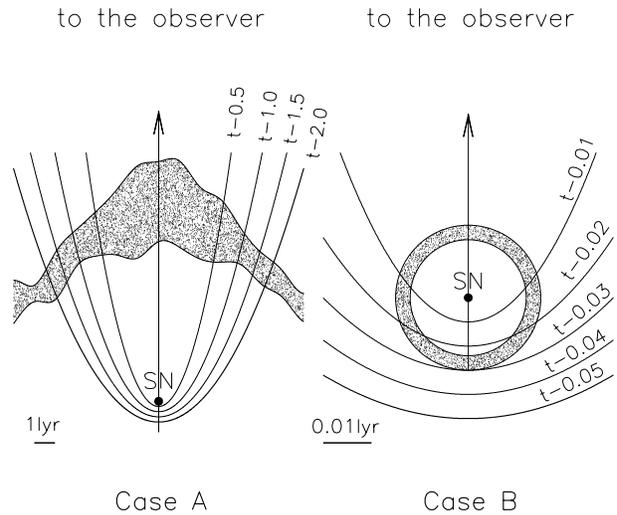}}
\caption{\label{fig:dustgeom}Schematic representation of dust and LE geometry for 
example cases A (large dust cloud, $t_c>$2 yr; left panel) and B
(circumstellar dust shell, $t_c=$0.015 yr; right panel) for a generic
observation time $t$ ($t\geq$2.0 yrs for Case A and $t\geq$0.05 yrs
for Case B). The parabolas represent the iso-delay surface
corresponding to some preceding epochs.}
\end{figure}

If we indicate with $C_{ext}(\lambda)$ the dust extinction cross
section, $\omega(\lambda)$ the dust albedo and $g(\lambda)$ the
forward scattering degree, in all those cases when there is one
single cloud with dust on the line of sight (like in the two examples
of Fig.~\ref{fig:dustgeom}), the formal solution of
equation~(\ref{eq:spectrum}) can be written as follows (see Paper I):

\begin{equation}
\label{eq:ss}
S_{LE}(\lambda,t) =   G[g(\lambda),t]\; \omega(\lambda) \;
C_{ext}(\lambda)\; e^{\tau_d(\lambda)}\; 
{\cal S}(\lambda,t)
\end{equation}

where $G[g(\lambda),t]$ is a time- and wavelength-dependent function
that includes dust and dust geometry properties and

\begin{equation}
\label{eq:integrated}
{\cal S}(\lambda,t)=\int_{t-t_c} ^t S(\lambda,t-t^\prime)\; dt^\prime
\end{equation}

is the time integrated SN spectrum, where $t_c$ is the time taken by
the generic iso-delay surface to completely cross the dust cloud. The
crossing time depends of course on the dust geometry and, if $t_c>t$,
the lower integration boundary of Equation~\ref{eq:integrated} has to
be replaced with 0.

When the SN is only mildly reddened ($\tau_d\ll$1), the resulting LE
spectrum is therefore proportional to the product between ${\cal
S}(\lambda,t)$ and the extinction cross section. When the optical
depth is larger ($\tau_d>0.3$), single scattering description must be
replaced by the more realistic {\it single scattering plus
attenuation} approximation (hereafter SSA; see Paper I), which holds
for $\tau \leq 1$. In this more refined schema, which takes into
account the LE self-absorption, equation~(\ref{eq:ss}) becomes

\begin{equation}
\label{eq:ssa}
S_{LE}(\lambda,t) = G[g(\lambda),t]\; \omega(\lambda) \; C_{ext}(\lambda)\; 
e^{\tau_d(\lambda)-\tau_{eff}(\lambda,t)}\; {\cal S}(\lambda,t)
\end{equation}

where $\tau_{eff}(\lambda,t)$ can be regarded as a weighted LE optical
depth at a given time $t$. In general, one has that $\tau_d \neq
\tau_{eff}$, as in the case sketched in Fig.~\ref{fig:dustgeom} (left panel),
where the SN suffers a larger extinction than the LE. In the cases
where the SN and the LE are affected roughly by the same extinction,
the two effects tend to compensate and the LE spectrum is again
proportional to the product between the observed time integrated SN
spectrum ${\cal S}$ and the extinction cross section $C_{ext}$.  These
formal expressions are useful to understand the principles of the
problem and will be used later on to explain the apparent peculiarity
reported by \cite{schmidt} in the extinction function (see
Section~\ref{sec:ext}).  Nevertheless, we note that the exact SSA
numerical solution is easy to implement in a code and it offers the
advantage of properly taking into account the attenuation effect and
the wavelength dependency of all involved quantities. While we refer
the reader to Paper I for more details, here we will spend some words
to discuss one important geometrical effect, which is related to
forward scattering and was only marginally mentioned in our previous
paper.

\begin{figure}
\centering
\resizebox{\hsize}{!}{\includegraphics{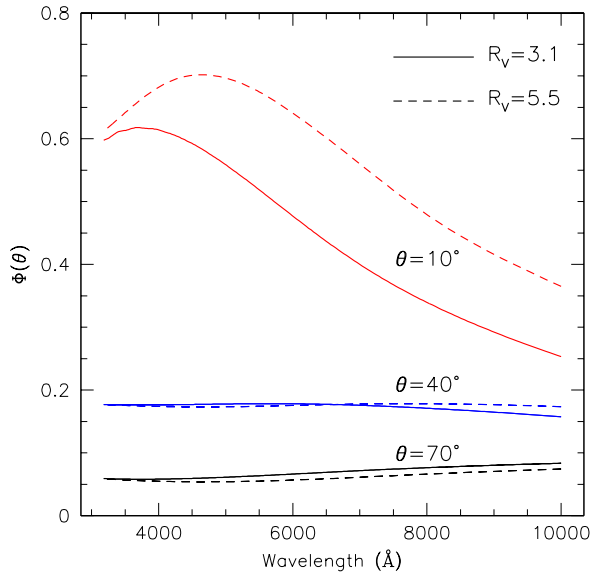}}
\caption{\label{fig:gtest}The Henyey-Greenstein scattering phase function 
$\Phi(\theta)$ computed for three different scattering angles and for
$R_V$=3.1 and $R_V$=5.5. The forward scattering degree function $g$ is from 
\citealt{draine}.}
\end{figure}

In fact, all LE calculations include the so called scattering phase
function $\Phi(\theta)$, which describes the probability distribution
of photon scattering angle $\theta$ (see for example
\citealt{chevalier}).  Usually, this function is parameterized,
following \cite{hg}, through the forward scattering degree
$g(\lambda)$.  To illustrate the behaviour of $\Phi(\theta)$, in
Figure~\ref{fig:gtest} we have plotted the results one obtains using
the Milky Way dust mixtures modeled by
\citet{draine} for $R_V$=3.1 (solid lines) and $R_V$=5.5 (dashed
lines) and for three values of the scattering angle. As one can see,
the wavelength dependency is much stronger for lower values of
$\theta$, with the scattering being more efficient in the blue than in
the red. Therefore, when the dust is located far in front of the SN,
which implies a small average scattering angle, the LE spectrum is
expected to be bluer than in the case where the same dust, with the
same geometry, is placed close to the SN.  As a consequence, the same
dust cloud can produce the same extinction to the SN (i.e. the same
$\tau_d$) but different LE colours, according to its position along
the line of sight. The simulations show that the $(B-V)$ colour of the
LE can change by $\sim$0.15 mag from a configuration where the SN is
embedded in the dust cloud to one where the same cloud is placed at
more than 200 lyr in front of the SN.

\section{Spectral Library}
\label{sec:speclib}

The LE spectral modeling relies on the availability of input SN
spectra. For this purpose one needs to build a spectral library with a
good phase coverage and time sampling, especially around maximum
light. This can be understood inspecting the cumulative functions for
the template light curves we discussed in Paper I. These were
constructed using the data of SNe 1992A and 1994D, two standard, well
studied and low reddening Type Ia (\citealt{kirshner};
\citealt{patat}). For $B$ and $V$ passbands, where the LE is expected
to be brighter, it turns out that about 90\% of the radiation is
emitted in 0.2 yr, while the radiation emitted in the first 10 days
accounts for about 5\% of total. Therefore, the spectral library
should include spectra from about a week before maximum to 2-3 months
past maximum. Spectra at later epochs are needed only if one is going
to extend the calculations to correspondingly late phases for dust
geometries with $t>t_c$.

\begin{figure}
\centering
\resizebox{\hsize}{!}{\includegraphics{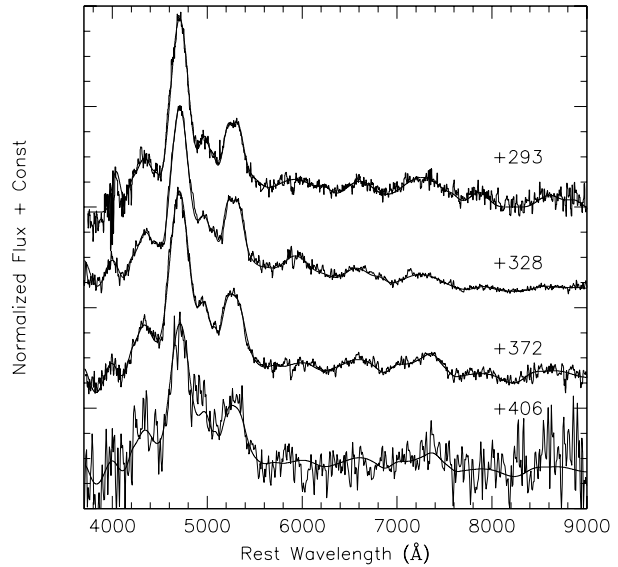}}
\caption{\label{fig:late}Late time spectra of SN~1992A (+328, +406) 
and SN~1994D (+293). The spectrum at +372 is an average of two spectra
of SN~1994D and 1992A at very similar phases. The solid curves in the
first three spectra trace the multi-Gaussian fittings. For the +406
spectrum, the solid line is the best fit to the +372 spectrum scaled
to match the observations.}
\end{figure}

We have extended the library used in Paper I, which included only data
from SN~1998aq \citep{branch}, adding early and late spectra from
SN~1992A (ESO Key-Programme on SNe) and SN~1994D
\citep{patat}, so that the data set ranges from $-$11 up to $+$406
days. Irrespective of the original flux calibration, the input spectra
have been re-calibrated to make their synthetic $V$ magnitude
consistent with the $V$ template light curve. As for the
late time spectra, we have used the available data of SNe 1992A and
1994D, which are presented in Figure~\ref{fig:late} and reach 406 days
past $B$ maximum light. As one can see, the spectra of the two objects
remain very similar, also at these late stages.  In order to remove
the noise in the data, we have used a multi-Gaussian fit to the
spectra (solid smooth lines). 

Due to the lack of data at later phases, in our calculations we have
assumed that after phase +406 there is no evolution and, from that
phase on, the SN spectrum is simply scaled according to the
exponential decay of the $V$ light curve. As a matter of fact, to our
knowledge the only Type Ia SNe with optical spectroscopy at phases
later than 450 days are SNe 1991T and 1998bu, both dominated by
evident LEs. Therefore, very little is known about the optical
spectral behaviour at these phases. Even though there are evidences of
deviations from the radioactive decay in the near-IR, as in the cases
of SN~1998bu \citep{jason} and SN~2002cx \citep{sollerman}, for the
optical domain things appear to be different. In fact, photometric
observations of the normal Ia SNe~1972E \citep{kirshner75}, 1992A
\citep{cappellaro} and 1996X \citep{salvo} at phases as late as 500
days indicate that the light curve follows the radioactive decay at
least until these epochs.

\begin{figure}
\centering
\resizebox{\hsize}{!}{\includegraphics{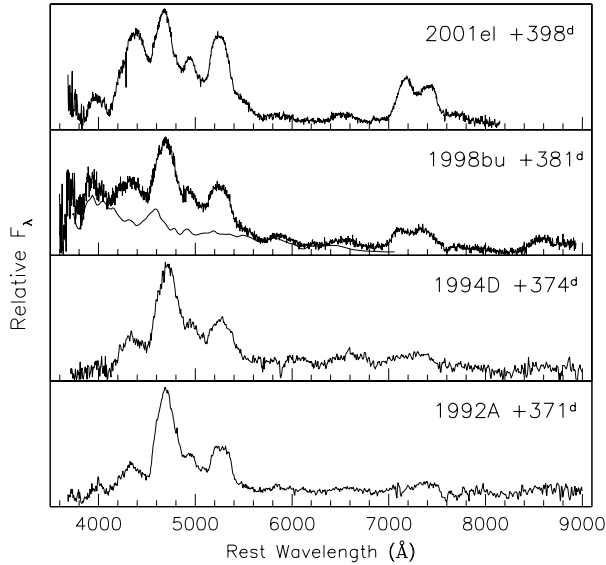}}
\caption{\label{fig:compialate}Comparison between four normal Type Ia at late 
phases: SN1992A (ESO-Key Programme Archive), SN1994D (ESO-Key
Programme Archive), SN1998bu \citep{jason} and SN2001el
\citep{mattila}. Epochs are computed from the $B$ maximum light. The thin 
curve in the third panel from bottom is a synthetic LE spectrum (see text).
SNe 1998bu and 2001el spectra were corrected for reddening using a canonical 
$R_V$=3.1 extinction law with $A_V$=0.96 \citep{jha} and $A_V$=0.57 
\citep{krisciunas}, respectively.}
\end{figure}

Another fact that should be mentioned is that even objects which were
very similar at early phases, do show differences at late phases. This
is clearly illustrated in Figure~\ref{fig:compialate}, where we have compared
the spectra of four Type Ia at about one year past maximum light: 1992A
(\dm15=1.47$\pm$0.05, \citealt{phillips}), 1994D (\dm15=1.32$\pm$0.05, 
\citealt{patat}), 1998bu (\dm15=1.01$\pm$0.05, \citealt{suntzeff}) and
2001el (\dm15=1.13$\pm$0.04, \citealt{krisciunas}). Despite the fact
that all these SNe have shown very similar spectra at maximum light
(\citealt{kirshner}, \citealt{patat}, \citealt{hernandez},
\citealt{wang}), at late phases some differences are visible. The most
pronounced one is the variable intensity of the bump at about 7300
\AA, usually attributed to a blend of [FeII] lines \citep{bowers}, which is
almost invisible in SN~1992A and well developed in SN~2001el. Other
discrepancies are observed also in the blue, where the [FeII] and
[FeIII] bumps appear to show different ratios in the various objects.
In this respect, it is worthwhile noting that the spectrum of
SN~1998bu is affected by a LE (see \citealt{cappellaro};
\citealt{jason}), which dominates below 4500\AA\/ (see 
Figure~\ref{fig:compialate}, third panel, thin curve).

The existence of these differences must be kept in mind, since the
intrinsic properties might be mistaken with the onset of a LE, even
when this is not the case.

\begin{figure}
\centering
\resizebox{\hsize}{!}{\includegraphics{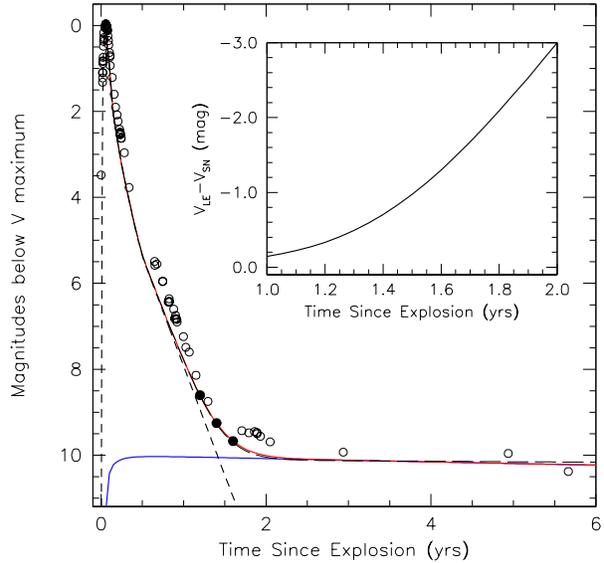}}
\caption{\label{fig:lcsheet} Light curve calculated for a
perpendicular sheet with a thickness $\Delta R$=50 lyr placed in front
of the SN with $D$=200 lyr and $n_H$=5 cm$^{-3}$ ($\tau_d(V)$=0.12,
solid line) and $D$=800 lyr and $n_H$=40 cm$^{-3}$ ($\tau_d(V)$=0.99,
long-dashed line). The short-dashed line traces the exponential decay
of the template light curve, the circles are the observed data for
SN~1991T \citep{schmidt,sparks} and filled dots mark the phases 1.2,
1.4 and 1.6 yr. The insert shows the deviation of the measured
magnitude from the radioactive decay tail for the first geometry.}
\end{figure}

\section{Effects on SN spectra and light curves}
\label{sec:effects}

Using the library we have presented here and the SSA approximation (or
the Monte Carlo approach when required) discussed in Paper I, we have
computed the LE spectra and analyzed the effects these have on objects
like SNe~1992A and 1994D under two very different conditions,
i.e. when the dust cloud is very large ($t_c\geq t$) and in the
opposite case when its size is very small ($t_c\ll t$).

\subsection{Case A: large clouds ($t_c\geq t)$}
\label{sec:fardust}

To illustrate the LE effects in the large cloud case (see
Fig.~\ref{fig:dustgeom}, left panel), we run a series of calculations
using the SSA approximation and a simple dust geometry, placing a dust
sheet perpendicular to the line of sight at $D$=200 lyr in front of
the SN, with a thickness $\Delta R$=50 lyr and a particle density
$n_H$=5 cm$^{-3}$. This configuration, which implies $\tau_d(V)$=0.12
(or $A_V$=0.13) produces a light curve which is qualitatively similar
to what was observed in 1991T
\citep{schmidt,sparks}, as shown in Figure~\ref{fig:lcsheet}. We
remark that we did not make any special attempt to fit the SN~1991T
data since our purpose here is to show the general effects in a
realistic situation. In this respect, it is also worth mentioning that
the light curve of this slow declining object and its spectra at
maximum where rather different from those we have used for our
calculations (see for example the broad light curve peak shown by
1991T in comparison to our template in Figure~\ref{fig:lcsheet}). A
more detailed analysis of the SNe 1991T and 1998bu cases is going to be
presented in a separate paper.

As shown in Figure~\ref{fig:lcsheet}, the light curve starts to
deviate from the ordinary exponential decay (dashed line) at about one
year after the explosion, to settle down on a very slow decline,
totally driven by the LE, at about two years past the
explosion. Between these two extremes, the observed spectrum is
expected to show a gradual transition from that typical of a Ia at
late phases to that of a pure LE. %This is clearly illustrated in
%Figure~\ref{fig:surfspec}, where we present the evolution between 1
%and 2 yr after the explosion, with a step of 0.1 yr.

\begin{figure}
\centering
\resizebox{\hsize}{!}{\includegraphics{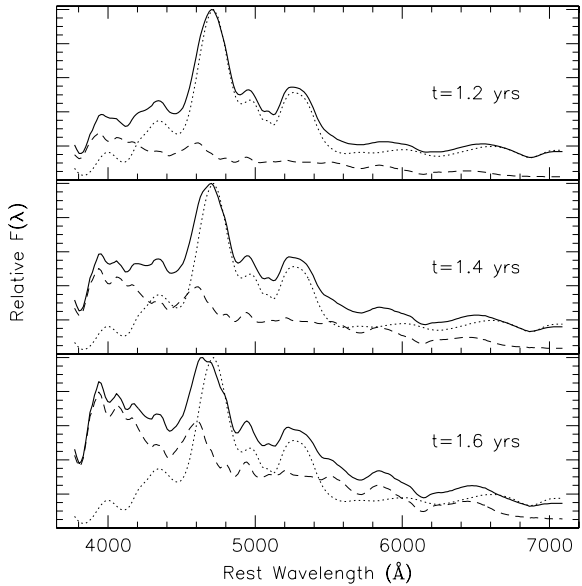}}
\caption{\label{fig:evol} Evolution of the global spectrum (SN+LE, thick curve) at 
three different epochs for the model described in the text. The dashed
line traces the LE spectrum while the dotted line is the SN spectrum
reddened by $A_V$=0.13. For presentation, SN and global spectra have been
normalized to their maxima.}
\end{figure}

The amount of the LE contamination can be judged in
Figure~\ref{fig:evol}, where we have plotted the results for $t$=1.2,
1.4 and 1.6 yr from the explosion. Already at the first epoch, when
the total $V$ luminosity (SN+LE) is only 0.2 mags brighter than that
expected for the radioactive decay, the resulting spectrum is sensibly
different from that of a Ia at these phases, especially in the
blue. Due to the peak present in the LE spectrum at about 4600\AA, the
most prominent feature of the intrinsic nebular SN spectrum at
4700\AA, normally attributed to [FeIII], appears to be broadened and
the region between 4000 and 4500 \AA\/ much bluer than in an
unaffected Ia.  This is actually quite similar to what was seen in
SN~1998bu \citep{jason} in the spectrum obtained 381 days past $B$
maximum (see also Figure~\ref{fig:compialate} here), where the bump at
about 4000
\AA\/ is unusually bright.

\label{sec:closedust}
\begin{figure}
\centering
\resizebox{\hsize}{!}{\includegraphics{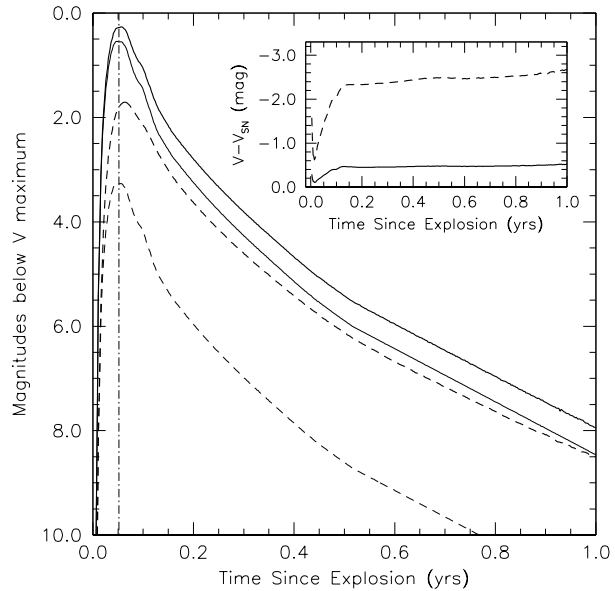}}
\caption{\label{fig:lcwind} Light curve calculated with the MC code 
for the $r^{-2}$ wind with $R_0$=0.01 lyr. The curves trace the global
(SN+LE, thick) and extincted SN (thin) light curves for $\tau_d(V)$=0.5
(solid) and $\tau_d(V)$=3.0 (dashed). The insert shows the deviation
of the global light curve (SN+LE) from the SN light curve for the two optical
depths. As a guide for the eye, the vertical dashed-dotted line marks
the position of $V$ maximum.}
\end{figure}

The consequences of a LE in the red are milder, and tend to emerge
only at later phases. The dominant [FeIII] bump gets gradually fainter
compared to the other emerging features and the colour becomes bluer
and bluer, reaching $(B-V)\simeq-$0.2 when the spectrum is totally
dominated by the LE, around 2 yr after the explosion. During the
transition phase a distinguishing feature is also the appearance of
CaII H\&K lines, reflecting their strong presence in maximum light
spectra.
%From these results it is clear that the presence of a LE might go
%unnoticed only in poorly studied objects, since its effects become
%well detectable, both in the spectra and in the light curve. LE
%effects might be misidentified with genuine SN characteristics only
%for objects with sparse data, not covering the LE dominated phases.

The outcome can be different if the dust optical depth is increased,
mainly due to the LE self absorption (see Paper I). To illustrate this
fact, we have run a series of calculations placing the dust sheet at
$D$=800 lyr and increasing its optical depth to $\tau_d(V)$ to about 1
($n_H$=40 cm$^{-3}$). The distance was properly tuned to produce a LE
with the same observed magnitude as the one discussed earlier (see
Figure~\ref{fig:lcsheet}, long-dashed line). In this case the LE
spectrum is much redder, having $(B-V)\sim$+0.15. As a matter of fact,
both the SN and the LE undergo roughly the same reddening, which in
this case is $E(B-V)\simeq$0.35.

In general, for higher extinctions the SSA approximation becomes
insufficient and a more complicated multiple scattering
approach has to be performed (see Paper I). 

%\begin{figure*}
%\centering
%\resizebox{\hsize}{!}{\includegraphics[angle=-90]{surfspec.eps}}
%\caption{\label{fig:surfspec}Evolution of the observed spectrum between 
%1 and 2 yr after the explosion. The spectra were calculated for a
%perpendicular sheet placed at $D$=200 lyr in front of the SN, with a
%thickness $\Delta R$=50 lyr and particle density $n_H$=5 cm$^{-3}$
%($\tau_d(V)$=0.12). For presentation each spectrum has been normalized
%to its maximum in the plotted range.}
%\end{figure*}

Before moving to the next section we like to make a consideration on
the late spectrum shown by SN~2001el. Besides the intense bump at
about 7500\AA, the region between 4200 and 5400\AA\/ appears to be
rather blue. In particular, the bump at 4400\AA\/ is much more intense
than in any of the other objects presented in
Figure~\ref{fig:compialate}. The synthetic colour of the observed
spectrum is $(B-V)$=0.25, and it reduces to $(B-V)$=0.07 after
applying the extinction reported by
\citet{krisciunas} for $R_V$=3.1. This value is unusually
blue when compared to the ones derived from the spectra of SN~1994D
(0.38) and SN~1992A (0.34).

\begin{figure}
\centering
\resizebox{\hsize}{!}{\includegraphics{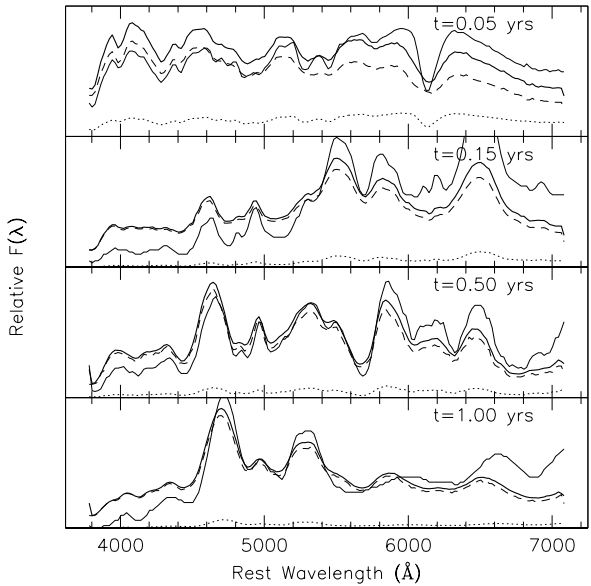}}
\caption{\label{fig:mcevol2} Same as Figure~\ref{fig:mcevol1} for 
$\tau_d(V)$=3.0 ($A^0_V$=3.26).}
\end{figure}

At a first glance this might be interpreted as the signature of an
underlying LE, but there are two quite important aspects that need to
be considered. The first is that the synthetic magnitude deduced from
the observed spectrum is $V$=21.24, which is in perfect agreement with
the standard luminosity decline. Secondly, the spectrum drops quite
fast below 4300\AA, contrarily to what happens for instance to 1998bu
(see Figure~\ref{fig:compialate}). The simulations show that changing
the optical depth of the region which is responsible for the
hypothetical LE does not allow to produce the drop which is seen in
the blue.  The conclusion is that the late time spectral appearance of
SN~2001el is not caused by a LE of the type considered in this
section.

In general, when the dust region is sufficiently extended in the
direction perpendicular to the line of sight, the dust system
integrates the emitted SN spectra over the whole time range from the
explosion to the epoch of the observation ($t_c\geq t$), so that the
resulting LE spectrum is very different from that of the SN at any
given epoch. Therefore, it is very easy to detect its emergence, that
can be difficultly mistaken with intrinsic features. An increase in
the dust density would simply anticipate the time when the LE emerges,
change the colour of the LE itself and cause a larger extinction on
the SN itself, but the LE characteristic spectral features will still
be very well pronounced.

\subsection{Case B: small clouds ($t_c\ll t$)}

\begin{figure}
\centering
\resizebox{\hsize}{!}{\includegraphics{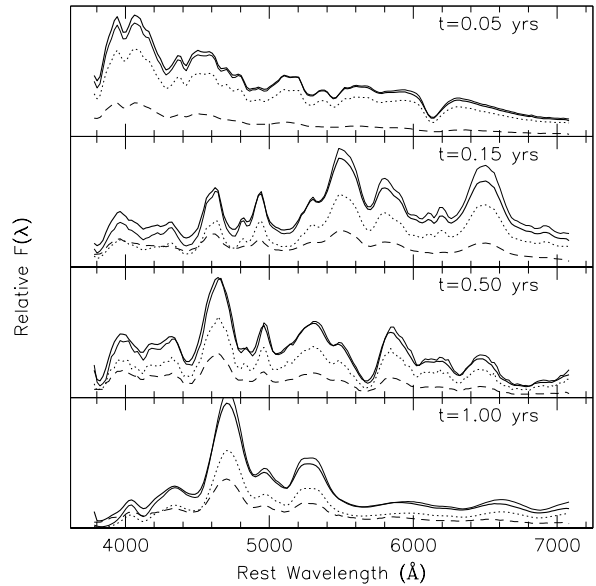}}
\caption{\label{fig:mcevol1} Evolution of the global spectrum (SN+LE, thick curve) at 
four different epochs for the wind model with $R_0$=0.01 lyr and
$\tau_d(V)$=0.5. The dashed line traces the LE spectrum while the
dotted line is the SN spectrum reddened by $A^0_V$=0.54. For
comparison, the latter spectrum has also been scaled to match the
integrated flux of the global spectrum (SN+LE, thin solid line).}
\end{figure}

\tabcolsep 1.6mm
\begin{table*}
\caption{\label{tab:models} LE effects on a $r^{-2}$ dust distribution for 
three different values of the inner wind boundary $R_0$ (see
text). All values are in mag, with the exception of $\gamma(B)$
(mag/100$^d$) and $\Delta t_{max}$ (days). The number under the labels
of columns 4, 5 and 6 indicate the reference values for the input
light curve. The meaning of each column is explained in the
footnotes.}
\begin{tabular}{ccccccccccccc}
\hline
$\tau_d(V)$ & $A^0_V$ & $E^0(B-V)$ & $\Delta m_{15}(B)$ & $\gamma(B)$ & $\Delta m_{300}(B)$ & 
$\Delta t_{max}$ & $\Delta V_{max}$ & $\Delta (B-V)_{max}$ & 
$E(B-V)_{max}$ & $A_V$ & $A^\prime_V$ & $\Delta M_V$ \\ 
            &         &            & [1.45]             & [1.40]      & [6.98]  &
               &       &              &              \\               
\hline
\multicolumn{13}{c}{$R_0$=0.1 lyr}\\
\hline
0.1 & 0.11 & 0.04 & 1.36 & 1.29 & 6.70 & 0.0 &$-$0.02 &$-$0.01 & 0.03 & 0.09 & 0.08 & +0.01\\
0.5 & 0.54 & 0.18 & 1.16 & 1.10 & 5.91 & 1.2 &$-$0.10 &$-$0.05 & 0.13 & 0.44 & 0.40 & +0.04\\
1.0 & 1.09 & 0.35 & 0.95 & 1.01 & 5.23 & 2.6 &$-$0.24 &$-$0.07 & 0.29 & 0.84 & 0.88 &$-$0.04\\
2.0 & 2.17 & 0.70 & 0.55 & 0.93 & 4.29 & 3.4 &$-$0.48 &$-$0.16 & 0.54 & 1.69 & 1.68 & +0.01\\
3.0 & 3.26 & 1.05 & 0.30 & 0.89 & 3.65 & 4.8 &$-$0.75 &$-$0.25 & 0.80 & 2.52 & 2.48 & +0.04\\
\multicolumn{13}{c}{}\\
\multicolumn{13}{c}{$R_0$=0.01 lyr}\\
\hline
0.1 & 0.11 & 0.04 & 1.38 & 1.41 & 6.94 & 0.0 &$-$0.05 &$-$0.02 & 0.02 &	0.05 & 0.05 & \phantom{$-$}0.00\\
0.5 & 0.54 & 0.18 & 1.30 & 1.43 & 6.79 & 1.2 &$-$0.27 &$-$0.09 & 0.09 &	0.28 & 0.28 & \phantom{$-$}0.00\\
1.0 & 1.09 & 0.35 & 1.24 & 1.43 & 6.63 & 2.6 &$-$0.56 &$-$0.14 & 0.21 &	0.53 & 0.64 &$-$0.11\\
2.0 & 2.17 & 0.70 & 0.97 & 1.37 & 6.31 & 3.2 &$-$1.07 &$-$0.28 & 0.42 &	1.10 & 1.29 &$-$0.19\\
3.0 & 3.26 & 1.05 & 0.86 & 1.33 & 6.04 & 4.4 &$-$1.56 &$-$0.41 & 0.64 &	1.70 & 1.99 &$-$0.29\\
\multicolumn{13}{c}{}\\
\multicolumn{13}{c}{$R_0$=0.005 lyr}\\
\hline
0.1 & 0.11 & 0.04 & 1.41 & 1.41 & 6.96 & 0.0 &$-$0.06 &$-$0.02 & 0.02 &	0.00 & 0.05 & \phantom{$-$}0.00\\
0.5 & 0.54 & 0.18 & 1.39 & 1.45 & 6.88 & 1.1 &$-$0.30 &$-$0.09 & 0.08 &	0.24 & 0.25 &$-$0.01\\
1.0 & 1.09 & 0.35 & 1.29 & 1.48 & 6.78 & 1.5 &$-$0.60 &$-$0.17 & 0.18 &	0.49 & 0.55 &$-$0.06\\
2.0 & 2.17 & 0.70 & 1.19 & 1.49 & 6.60 & 2.8 &$-$1.17 &$-$0.30 & 0.40 &	1.00 & 1.23 &$-$0.23\\
3.0 & 3.26 & 1.05 & 1.03 & 1.47 & 6.46 & 3.1 &$-$1.69 &$-$0.44 & 0.61 &	1.57 & 1.90 &$-$0.33\\
\hline
\multicolumn{13}{l}{$A^0_V$=1.086$\tau_d(V)$; $E^0(B-V)$=$A_V^0$/3.1; $\Delta m_{15}(B)$=$B_{15}-B_{max}$;
 $\gamma(B)$=$(dB/dt)_{t>180}$;  $\Delta m_{300}(B)$=$B_{300}-B_{max}$; $\Delta t_{max}$=$t_{max}-t^0_{max}$;}\\
\multicolumn{13}{l}{$\Delta V_{max}$=$V_{max}-V_{max}^0+A^0_V$; $\Delta (B-V)_{max}$=$(B-V)_{max}-(B-V)^0_{max}+E^0(B-V)$;
$E(B-V)$=$E^0(B-V)+\Delta (B-V)_{max}$;}\\
\multicolumn{13}{l}{$A_V$=$A_V^0+\Delta V_{max}$; $A^\prime_V$=3.1$E(B-V)$; $\Delta M_V$=$A_V-A^\prime_V$.}\\
\end{tabular}
\end{table*}

More subtle is the case when the dust system characteristic dimension
is significantly smaller than $ct$ (see Fig.~\ref{fig:dustgeom}, right
panel). In fact, under these conditions, the dust system integrates
the SN input spectra over times which are much shorter than the time
elapsed from the explosion (i.e. $t_c\ll t$) and the resulting LE
spectrum is much closer to the SN spectrum, making it very difficult
to detect it on the basis of the spectral appearance alone. Of course,
in order to have this effect, the dimension of the dust cloud must be
indeed small. For example, for a spherical dust cloud with radius $R$
and centered on the SN it is easy to show that $t_c=2R/c$. The
simulations show that in order to have a LE spectrum which is very
similar to that of the SN at all epochs, one needs to have $t_c
\leq$0.2 lyr, which implies $R\leq$0.1 lyr ($\sim$10$^{17}$
cm)\footnote{If one is considering only phases as late as one year,
due to the slow spectral evolution the cloud size can be larger and
still produce a LE spectrum very similar to that of the SN.}. Since
these dimensions are much smaller than those of typical interstellar
dust clouds, the most natural scenario to be considered is the one of
circumstellar material. As done by \cite{chevalier},
\cite{paperI} and \cite{wang05}, we will consider here the simplest case, 
i.e. that of a spherically symmetric stellar wind in which the density
is given by $n(r)=n_0 (R_0/r)^2$, where $R_0$ is the inner boundary of
the wind, within which the density is supposed to be null. Also, we
assume that the wind has an outer boundary, defined by $R_1>R_0$. Under
these circumstances, the optical depth along the line of sight is
simply $\tau_d=C_{ext}\; n_o\; R_0 \; (1-R_0/R_1)$, from which it is
clear that if $R_1\gg R_0$ the value of $\tau_d$ is practically
independent from $R_1$. As we have shown in Paper I (cf. Fig.~6) this
geometry, compared to the others we had explored, is able to produce
quite bright LEs with a fast declining light curve.

To illustrate the effects of such a dust geometry, we have run a
series of calculations using the MC code we have described in Paper
I\footnote{Proper multiple scattering treatment is mandatory in all
cases where the dust is very close to the SN, especially when its optical
depth is larger than 1 (see Paper I)}. In all simulations we have used
the canonical Milky Way dust mixture with $R_V$=3.1
\citep{draine} and we have computed the LE $B$ and $V$ light curves
and optical spectra for $R_0$=0.1, 0.01 and 0.005 lyr, keeping
$R_1$=1.0 lyr and for different values of $\tau_d(V)$ (0.1, 0.5, 1.0,
2.0 and 3.0). Two examples of the resulting curves are presented in
Fig.~\ref{fig:lcwind} for $R_0$=0.01 lyr and $\tau_d(V)$=0.5, 3.0. The
main photometric parameters obtained from the global light curves
(SN+LE) are summarized in Table~\ref{tab:models}, where besides the
the optical depth and the implied extinction and colour excess
($\tau_d(V)$, $A^0_V$ and $E^0(B-V)$) we show the usual $\Delta
m_{15}(B)$, the slope at phases later than 6 months ($\gamma(B)$), the
magnitude jump between maximum and 300 days past maximum ($\Delta
m_{300}(B)$), the time shift in the $B$ maximum ($\Delta t_{max}$),
the differences in  $V$ and $(B-V)$ at maximum with respect to the
extincted SN ($\Delta V_{max}$, $\Delta (B-V)_{max}$), the colour
excess with respect to the unreddened SN at maximum ($E(B-V)_{max}$),
the apparent extinction obtained comparing with the intrinsic
magnitude ($A_V)$, the extinction derived from $E(B-V)_{max}$ assuming
$R_V$=3.1 ($A^\prime_V$) and the deviation of the absolute magnitude
at maximum with respect to the input value ($\Delta M_V$). For
comparison, the corresponding values for the template light curves
assumed by the models are as follows: $\Delta m_{15}(B)$=1.45 mag,
$\gamma(B)$=1.40 mag/100$^d$ and $\Delta m_{300}(B)$=6.98 mag.

There are several conclusions that can be drawn already inspecting
Table~\ref{tab:models}. First of all, as already qualitatively noticed
by \cite{wang05}, the resulting light curves become broader and the
maximum is attained later then in the pure SN case (cf. also
Fig.~\ref{fig:lcwind}). This has the clear effect of decreasing
$\Delta m_{15}(B)$ rather sensibly and the change increases with
$\tau_d(V)$ and $R_0$.  Moreover, the global colour gets bluer than
that of the purely extincted SN and this effect becomes stronger for
larger $\tau_d$ and smaller $R_0$. In principle, unless the values of
this parameters become too deviant from the {\it normal} ones (0.9
$\leq \Delta m_{15}(B)\leq$1.9, \citealt{phillips}), they could be
attributed to an intrinsic feature of the SN. Nevertheless, there are
other side effects that tend to betray the presence of an underlying
LE. In fact, the late time decay rate $\gamma(B)$ increases with
$\tau_d$ and $R_0$, while the magnitude jump $\Delta m_{300}(B)$ grows
with $R_0$ and decreases for increasing values of $\tau_d$.  In the
case of $R_0$=0.005 lyr ($\sim$5$\times$10$^{15}$ cm), the global
light curve around maximum light changes but its overall shape remains
similar to the template light curve (e.g. giving plausible values for
$\Delta m_{15}(B)$, $\Delta m_{300}(B)$ and so on). This has a very
interesting consequence, at least for $R_0\leq$0.01 lyr. In fact,
besides measuring a slower post-maximum decline rate, one would also
measure a colour which is significantly bluer and a global apparent
luminosity which is higher than those of the purely reddened SN. As
the simulations show, these two facts combined together would lead an
hypothetical observer deducing the reddening correction from the
global observed colour (and using a standard value for $R_V$) to derive an
absolute luminosity which is brighter than the real one by an amount
that can reach about 0.3 mag (see last column of
Tab.~\ref{tab:models}). In conclusion, one would attribute the smaller
value of $\Delta m_{15}(B)$ to a larger intrinsic luminosity, as in
the case of the classical Phillips relation
\citep{phillips}.  Of course this does not mean that the observed relation
is produced by dust in the vicinity of the SN, since $\Delta m_{15}$
correlates in fact with other features like spectral line ratios
\citep{nugent, benetti05, hachinger, bongard}. But certainly, the presence of hidden LEs
might introduce an additional spread in the measured photometric
parameters.

The effects on spectra are presented in Fig.~\ref{fig:mcevol1} and
\ref{fig:mcevol2}, where we show the time evolution in the $R_0$=0.01 lyr
case for $\tau_d(V)$0=0.5 and 3.0. In the low optical depth case the
effect is rather mild (the global spectrum is slightly bluer than
that of the reddened SN spectrum at all phases), while the LE
influence becomes much more relevant in the high optical depth
case. As a consequence of the increased LE contribution to the total
flux, the time integration becomes visible and this causes the
spectral features to show clear deviations from the intrinsic ones.
The most remarkable is the absorption profile of the Si~II 6355 \AA\/
line around maximum light ($t$=0.05 lyr), which is much broader. The
same behaviour is shown by all lines and at all epochs. A late phases,
the line peaks also appear blue-shifted with respect to the intrinsic
features. Moreover, as in the low density case, the global spectrum is
bluer than the reddened SN spectrum at all times.

\begin{figure}
\centering
\resizebox{\hsize}{!}{\includegraphics{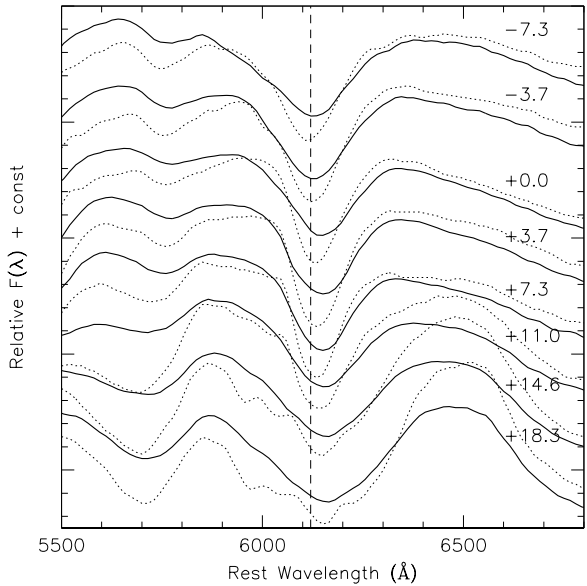}}
\caption{\label{fig:mctimeseq} Evolution of the Si~II 6355\AA\/ region for 
$R_0$=0.01 lyr and $\tau_d(V)$=3.0 between $-$7.3 to +18.3 days past
maximum light. The dotted curves trace the reddened SN spectra. For
presentation they have been scaled by a suitable amount in order to
match the global spectrum. The vertical dashed line marks the position
of the absorption minimum of Si~II in the first spectrum. In the MC
simulations the photons where collected with a resolution of $\Delta
\lambda$=15 \AA\/ and a time bin $\Delta T$=0.01 yrs.}
\end{figure}

In general, the strongest effect is expected around maximum light and
in the weeks that immediately follow it, i.e. when the spectral
evolution is fast compared with the LE integration time which, in the
case of $R_0$=0.01 lyr is of the order of 12 days\footnote{Due to the
$r^{-2}$ dust density distribution, this is larger than the 7.3 days
one would have in the case all the dust was located in a very thin
shell with a radius of 0.01 lyr.}. In the pre-maximum phase, the early
spectra contribute quite marginally to the global flux, due to the
very rapid rise in luminosity while, just after maximum, the SN flux
in the preceding days is actually higher than on the current epoch, so
that the contribution is indeed relevant.  This is clearly displayed
in Fig.~\ref{fig:mctimeseq}, where we present the calculated evolution
of the Si~II region between $t$=0.03 and $t$=0.10 yrs, which
correspond to $-$7.3 and +18.3 days past maximum light. On the first
epoch, the profile of the global spectrum is already broader than in
the input SN, but the deviation grows in the following days, reaching a
maximum during the two weeks past maximum light.  The position of the
absorption through minimum appears to be red-shifted with respect to
the input spectrum, producing a lower photospheric velocity
estimate. During the second week, also the emission peak of the line
is affected and it appears much broader, reflecting the higher
velocities reached in the preceding epochs.
These deviations, together with those shown by the light (and colour)
curves, would most probably allow one to recognize the presence of a
LE. Nevertheless, we must notice that this is true only for high
optical depths, while for $\tau_d(V)\leq$1, the effects on photometric
parameters and spectral features are rather mild (see
Table~\ref{tab:models} and Fig.~\ref{fig:mcevol1}), making the
detection of the LE certainly much more difficult. In those cases the
net effect of the LE would be to alter the colours and the overall flux
of the observed spectrum, with the consequences discussed by
\cite{wang05}, that we will analyze in the next section.

\section{Implications on extinction and extinction law}
\label{sec:ext}

If a SN is affected by the emergence of a LE, it is clear that each
time one tries to derive the amount of extinction or information about
the dust properties, the results will be in general misleading. As we
have done in the previous section, we distinguish here the two
opposite cases.

\subsection{Case A}
\label{sec:casea}

When \cite{schmidt} discovered the LE in SN~1991T, they noticed that,
in order to reproduce the observations, they had to correct the time
integrated SN spectrum $\cal S$ for the effects of scattering. These were
parameterized with a function of the type $\lambda^{-\alpha}$ and the
best fit was obtained using $\alpha$=2.  
%Taken as face value, this
%would be much larger than the canonical values typical of a $R_V$=3.1
%dust mixture. For example, a best fit to the model by \citet{draine}
%in the range 4000-9000\AA\/ gives $\alpha$=1.35 (see Paper I).
As we have seen in Section~\ref{sec:basic}, the correction one has to
apply to $\cal S$ includes three components: the first is related to
the light scattered by the dust into the observer's line of sight
(directly proportional to $C_{ext}$), the second accounts for the
reddening suffered by the SN (proportional to $e^{C_{ext}}$) while the
third describes the auto-absorption within the dust cloud that
generates the LE itself (proportional to $e^{-C_{ext}})$. Finally,
there is a mild wavelength dependency introduced by the albedo and a
more important contribution by the function $G(\lambda,t)$ (see
Section~\ref{sec:basic}).  A more convenient formulation can be
obtained passing from the extinction cross section $C_{ext}(\lambda)$
to the normalized extinction law $a(\lambda)$. If we pose

\begin{displaymath}
a(\lambda) = \frac{C_{ext}(\lambda)}{C_{ext}(V)}
\end{displaymath}

then the absorption (in magnitudes) can be expressed as
$A(\lambda)=a(\lambda) A_V$ and equation (\ref{eq:ssa}) can be
rewritten as

\begin{equation}
\label{eq:law1}
\frac{S_{LE}(\lambda,t)}{{\cal S}(\lambda,t)} \propto
G(\lambda,t) \; \omega(\lambda) \; a(\lambda) \;
e^{\frac{\kappa(V) \; a(\lambda)}{1.086}}
\end{equation}

where $\kappa(V)=A_d(V)-A_{eff}(V)$ is expressed in magnitudes. In a
first approximation, one can parameterize $a(\lambda)$ with a power
law of the type $\lambda^{-\beta}$. From equation (\ref{eq:law1}) it is
clear that a fitting to the ratio between $S_{LE}$ and ${\cal S}$
using a power law $\lambda^{-\alpha}$ would give a value for $\alpha$
which is not directly related to $\beta$, since other terms with
different wavelength dependencies enter its expression. In fact, using
our simulations for the perpendicular sheet with $D$=200 lyr we have
found that for the Milky Way dust mixture with $R_V$=3.1
\citep{draine}, the best fit values of $\alpha$ depend mostly on the
value of $\kappa(V)$, while the dependence on $\tau_{eff}$ is very
mild, at least in the validity range of the SSA approximation
($\tau_{eff}\leq$1). Typical values for $\alpha$ are 1.4, 1.7, 2.0 and
2.2 for $\kappa(V)$=$-$0.2, 0.0, 0.2 and 0.4 respectively.

\begin{figure}
\centering
\resizebox{\hsize}{!}{\includegraphics{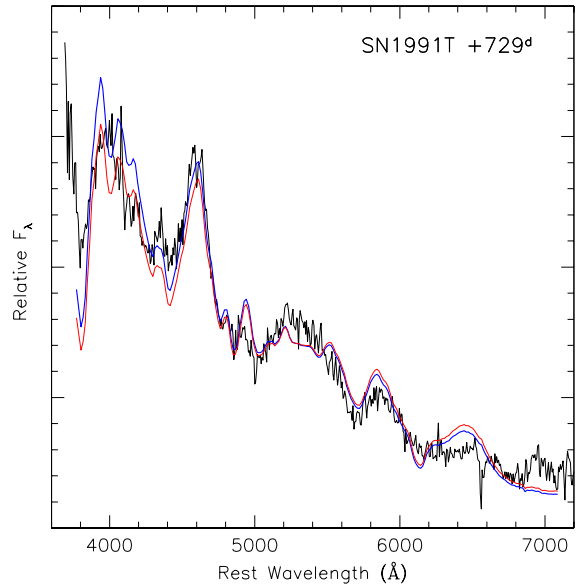}}
\caption{\label{fig:sn91t}Spectrum of SN~1991T at 729 days past $B$ maximum 
(ESO Key Programme). The spectrum was corrected for the Milky Way
reddening ($A_V$=0.07, \citealt{phillips}). The thick smooth curve is
a synthetic LE spectrum obtained with the spectral library described
in the text with $\tau_{eff}$=0.24. For comparison, a synthetic
spectrum with $\tau_{eff}$=0.42 is also shown (thin curve).}
\end{figure}

It is interesting to notice that the $\alpha$=2 value found by
\citet{schmidt} can be explained with a canonical extinction law and a
small difference between the extinction suffered by the LE and the SN,
i.e. only 0.2 mag. The SN was certainly reddened: \cite{phillips}
report in fact a total extinction $A_V$=0.53$\pm$0.17, 0.46 of which
is due to the host galaxy. If we assume $A_d$=0.46 (or equivalently
$\tau_d(V)$=0.42) we can therefore give an estimate to $A_{eff}$, which
turns out to be 0.26 (i.e. $\tau_{eff}(V)$=0.24). The corresponding
model is presented in Figure~\ref{fig:sn91t}. Given the fact that the
synthetic spectrum has been computed using the spectra of objects like
1994D and 1998aq which are quite different from SN~1991T, especially
during the maximum light phase, the match is reasonably good.
However, we must notice that the observed spectrum is also
compatible with $\kappa(V)$=0, i.e. with the SN and the LE
suffering the same amount of reddening (Figure~\ref{fig:sn91t}, thin
smooth curve).

Another example of a possibly similar situation is given by
SN~1998bu. \citet{cappellaro} have presented a LE spectrum at 670 days
past maximum light, with a $(B-V)$ colour close to 0, i.e. very
similar to that observed for SN~1991T
\citep{schmidt}. On the other hand, \citet{jha} have reported a total
extinction $A_V$=0.94 $\pm$0.15, while \cite{hernandez} found
$A_V$=1.0 $\pm $0.1.  Using the same geometrical configuration, the
best fit with the observed LE spectrum is obtained for
$\tau_{eff}(V)$=0.62, which implies $A_V$=0.67 (see
Figure~\ref{fig:sn98bu}). Given the fact that the estimated galactic
extinction in the direction of SN~1998bu is only $A_V$=0.08
\citep{schlegel}, this value is about 0.2 mag lower than the average value
reported by \citet{jha} and \citet{hernandez}.

In this respect it is interesting to note that, due to the geometrical
effect related to forward scattering that we have described in
Section~\ref{sec:basic}, if the dust is close to the SN, then a lower
dust optical depth would still produce a red LE spectrum.  This is not
because there is more reddening, but because under these circumstances
the scattering efficiency does not grow at shorter wavelengths as fast
as it does if the dust is located far way. This is illustrated in the
insert of Fig.~\ref{fig:sn98bu}, where we have plotted the results of
a simulation where the same perpendicular sheet is placed at distance
$D$=0. In this case the best fit is obtained for $\tau_{eff}(V)$=0.25,
which would imply a SN extinction $A_V$=0.27, which is significantly
lower than the values reported in the literature.

In conclusion, the information on the extinction and the extinction
law one derives from the observed LE spectrum is geometry dependent
and not univocally related to the dust properties.

\subsection{Case B}
\label{sec:caseb}

As we have seen in Sec.~\ref{sec:closedust}, the presence of a LE
produced by a dust geometry with $t_c\ll t$ is able to introduce
sensible luminosity and colour changes with mild effects on the other
photometric and spectroscopic characteristics, provided that
$\tau_d(V)\leq 1$. This, as \cite{wang05} has shown, can have an
interesting consequence on the properties one would deduce for the
dust. In fact, if we take the example of $R_0$=0.01 lyr and
$\tau_d(V)$=1.0, from Table~\ref{tab:models} we see that, due to the
presence of a LE, the global $V$ magnitude is 0.56 brighter than the
purely extincted SN, which is affected by an extinction of
$A^0_V$=1.09 mag.  Therefore, assuming a distance to the SN and using
the expected absolute luminosity, one would deduce $A_V$=0.53. Then,
comparing the colour at maximum with a reference value (which here we
assume to be the one of our template object), one would also deduce
$E(B-V)_{max}$=0.21 from which she would erroneously conclude that
$R_V\sim$2.5. As far as the photometric properties are concerned, such an
object would show normal $\Delta m_{15}(B)$ (1.24) and $\gamma(B)$
(1.43) and at late phases the slight over-luminosity (0.3 mag) would
probably go unnoticed due to the measurement errors. Therefore, at
least on the basis of those parameters, it would be very difficult to
recognize the presence of an underlying LE. 

\begin{figure}
\centering
\resizebox{\hsize}{!}{\includegraphics{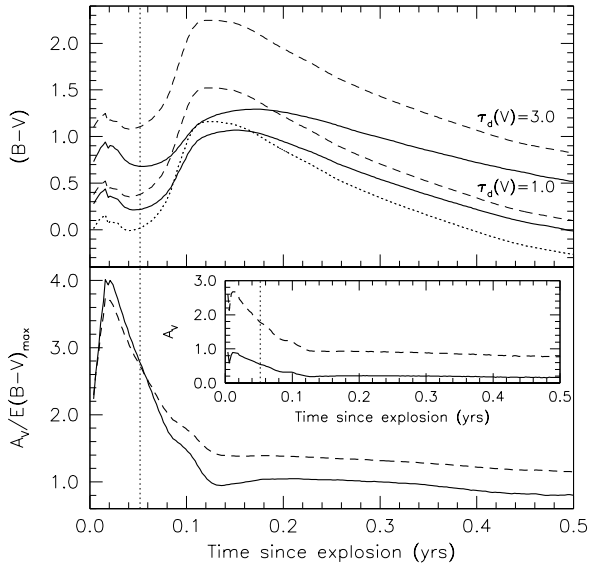}}
\caption{\label{fig:windmccol}Upper panel: colour evolution for the wind 
dust geometry with $R_0$=0.01 lyr for $\tau_d(V)$=1.0 (lower thick
curve) and $\tau_d(V)$=3.0 (upper thick curve). The dashed lines trace
the corresponding colour curve in the case of pure extinction and
$R_V$=3.1, while the dotted curve is the SN intrinsic colour. Lower
panel: ratio between $A_V=V-V^0_{SN}$ and $E(B-V)_{max}$ deduced
comparing the global colour with the intrinsic colour of the SN at
maximum for $\tau_d(V)$=1.0 (solid line) and 3.0 (dashed line). The
insert shows the behaviour of $A_V$ in the two cases. In all plots the
vertical dotted line marks the epoch of maximum light.}
\end{figure}

Since the difference between the global luminosity and that
of the pure SN is not constant in time and changes rather rapidly
around maximum light (see Fig.~\ref{fig:lcwind}, upper right insert),
the values of $R_V$ deduced in this way are expected to change with
time in a similar way. This is illustrated in
Fig.~\ref{fig:windmccol}, where we present the results of MC
simulations for $R_0$=0.01 lyr and two different values of the dust
optical depth. The total-to-selective absorption ratio changes in fact
very rapidly from the explosion to about 0.15 yrs, which correspond to
the maximum of the intrinsic SN ($B-V)$ colour curve (dotted
curve). After that epoch, the ratio remains almost constant. This
behaviour is qualitatively similar to the one found by \cite{wang05} (see
his Fig.~2) and the discrepancies are probably due to the different
treatments of multiple scattering\footnote{We reach results very similar to
those reported by \cite{wang05} for $\tau_d(V)$=0.5.}.

It is important to remark that the global colour curve is not obtained
by a simple rigid reddening of the unextincted SN colour curve. On the
contrary, the difference between bluest and reddest colour is smaller
and the tail at phases later than 0.15 yrs ($\sim$55 days) is less
steep. The effect becomes more pronounced for higher optical depths.
In other words, once the observed global colour curve is corrected for
the colour excess deduced at maximum, this is not going to match a
reference colour curve and the late behaviour would clearly differ from
the homogeneous trend found by \cite{lira}. We reckon that this colour
effect, shown also by the semi-analytical calculation of \cite{wang05} (see his
Fig.~3, lower panel) is a good diagnostic to unveil the presence of a
hidden LE.

At this point we must emphasize that, in all practical cases, the
colour excess is deduced from the data using the colour curve. On the
contrary, the extinction is usually derived from this value after
assuming $R_V$ and then applied to the observed magnitude to obtain
the absolute luminosity of the object (see for example
\citealt{benetti}). Following this procedure in the case of our
previous example would lead to a result which is actually not far from
the real one. In fact, if one adopts $E(B-V)$=0.21 and assumes
$R_V$=3.1, this would produce an absolute magnitude which is only 0.11
mag brighter than the real value (see Table~\ref{tab:models}, last
column). Remarkably, this is of the same order of the correction foreseen by the
Phillips relation \citep{phillips} for the given $\Delta
m_{15}(B)$. As we had already noticed in Sec.~\ref{sec:closedust},
this effect is most likely going to add a spread to the observed
luminosity-decline rate relation.

\begin{figure}
\centering
\resizebox{\hsize}{!}{\includegraphics{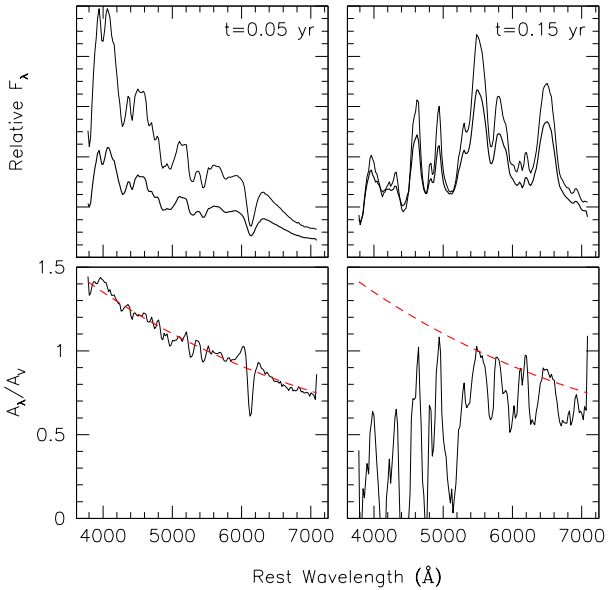}}
\caption{\label{fig:testext} Upper panels: comparison between the global 
spectrum (thick curve) and the intrinsic SN spectrum (thin
curve). Lower panels: comparison between intrinsic ($R_V$=3.1, dashed
curve) and derived (solid curve) normalized extinction laws. The
simulations were performed using $R_0$=0.01 lyr and $\tau_d(V)$=1.0.}
\end{figure}

Another way of deriving information on the extinction law $A(\lambda)$
is to compare the observed spectrum $S(\lambda)$ with a reference
unreddened SN spectrum $S^0(\lambda)$ obtained at the same phase and
corrected to the same distance (see for instance \citealt{nancy}). If
the LE effects were negligible, one could in fact write
$A(\lambda)=1.086\; \ln[S^0(\lambda)/S(\lambda)]$ and hence derive
directly the extinction wavelength dependency. Of course, when
$S(\lambda)$ contains a significant contribution from the LE the
result is going to be misleading. This is clearly illustrated in
Fig.~\ref{fig:testext}, where we present the results obtained from our
Monte Carlo simulations at two different epochs ($t$=0.05 and 0.15
yrs) for the case $\tau(V)$=1.0 and $R_0$=0.01 lyr. For the first
epoch, which approximately corresponds to the $B$ maximum light, the
colour of the global spectrum is clearly redder than that of the
unreddened SN ($E(B-V)\sim$0.2, see also Fig.~\ref{fig:windmccol}) and
the derived extinction law is just slightly steeper than the input
one, thus giving a smaller $R_V$. With the exception of the SiII
6355\AA\/ feature, the derived curve is reasonably smooth.  Totally
different is the case at $t$=0.15, which roughly corresponds to the
maximum of the $(B-V)$ colour curve. At this phase, in fact, the
colours of the global spectrum and the unreddened SN are similar (see
also Fig.~\ref{fig:windmccol}) and the derived extinction law is
dominated by the differences displayed by the two spectra due to the
time integration effects. Of course, in a real case, this could be
interpreted as the presence of a LE but also with a deviation from the
standard spectral evolution. Irrespective of the physical reason, the
apparent evolution shown by the derived extinction law is a clear
diagnostic signaling that additional effects are present and that
$R_V$ can not be estimated with this method.  Given these facts, a
detailed analysis of extincted Ia should be able to clarify whether
this mechanism is indeed at work and whether it can explain the low
values of $R_V$ derived by several authors.

\section{Discussion}
\label{sec:discuss}

All known cases of LEs in Type Ia SNe have revealed themselves through
the gradual appearance of an additive flux, during the nebular phase,
in the spectral region bluewards of 4500\AA, where one can easily
recognize the imprints of the time integrated SN spectrum. Therefore,
they are instances of the Case A we have described in
Sec.~\ref{sec:fardust}. The spectrum discussed by
\cite{jason} is a very good example of this behaviour (see also 
Figure~\ref{fig:compialate} here).  Moreover, due to the close
wavelength coincidence of the dominating [Fe III] nebular feature with
the most prominent bump of a LE spectrum, the former line is
broadened.  Besides this, a distinguishing spectral feature of a LE is
the clear presence of the Ca II H\&K lines which, though, at the
beginning of the transition phase can be masked by the presence of a
similar structure observed in pure nebular spectra at about 4000 \AA.

\begin{figure*}
\centering
\resizebox{\hsize}{!}{\includegraphics{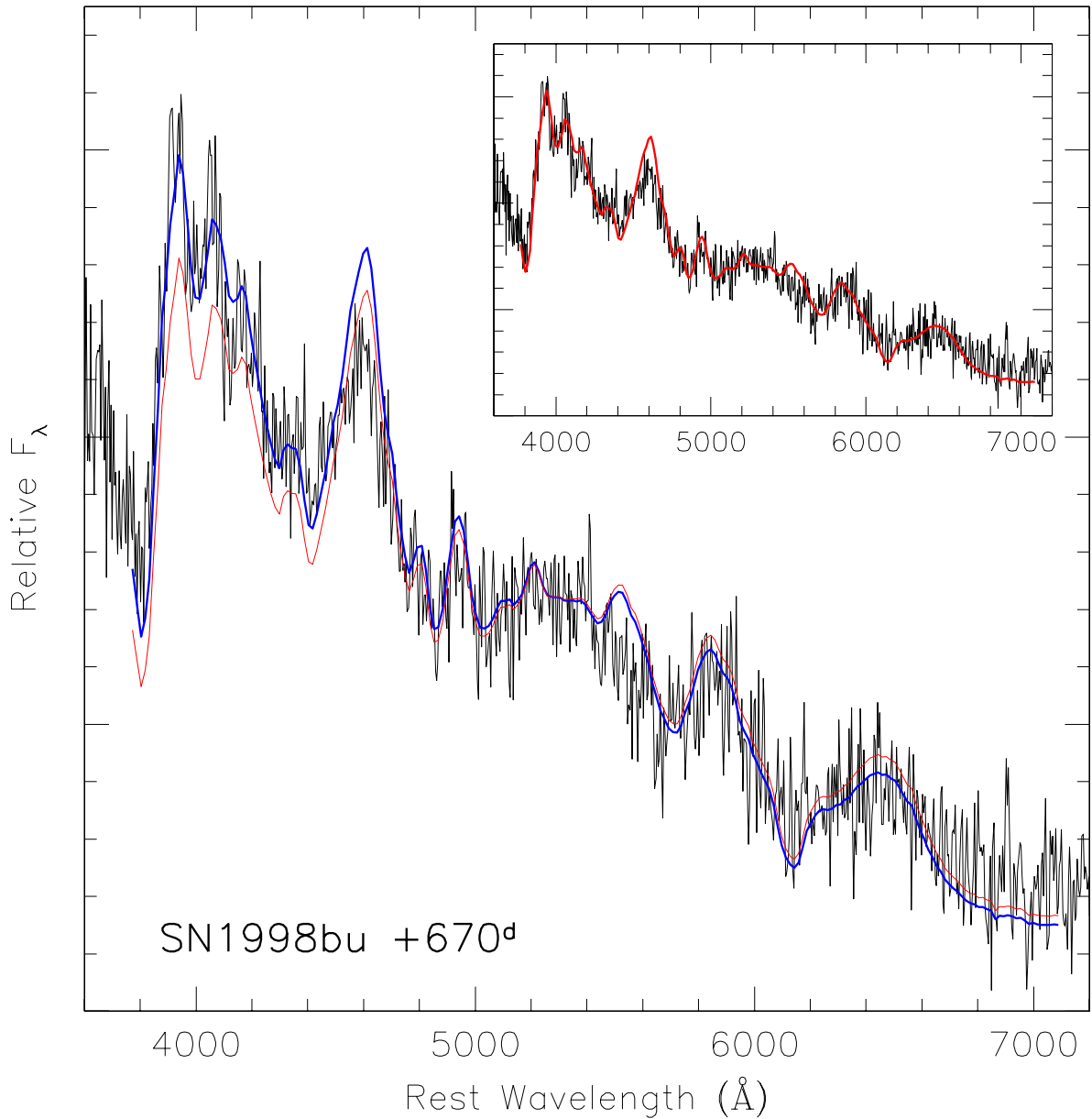}}
\caption{\label{fig:sn98bu}Spectrum of SN~1998bu at 670 days past $B$ maximum 
\citep{cappellaro}. The spectrum was corrected
for the Milky Way reddening ($A_V$=0.08, \citealt{jha}). The smooth
curves are synthetic LE spectra with $\tau_{eff}$=0.62 (thick) and
$\tau_{eff}$=0.90 (thin) respectively. The upper right insert shows
the results for the same dust sheet placed at $D$=0 and $\tau_{eff}$=0.25
($A_V$=0.27).}
\end{figure*}

In general it is fairly easy to recognize the Case A LE contamination
on the basis of the late spectra alone, especially if the optical
depth is not too high, so that its spectrum is definitely blue, with
$(B-V)\leq$0. In cases where the LE forms in regions of high dust
density, its spectrum can turn red without necessarily having the SN
spectrum reddened. In this case, the LE contamination is a bit more
subtle and tends to leave unchanged the overall colour of the observed
spectrum. However, the characteristic LE bumps at about 5900\AA\/ and
6500\AA \/ (see Figure~\ref{fig:evol}) appear quite clearly.

An interesting fact that emerges from the analysis of the two known
cases of SNe 1991T and 1998bu when compared to our synthetic LE
spectra is that clear differences are visible. As far as 1991T is
concerned, its LE dominated spectrum taken at 729 days past maximum
(cfr. Figure~\ref{fig:sn91t}) shows interesting features, which are
directly related to its peculiar appearance during the maximum phase
(see for instance \citealt{filippenko}). In particular, we notice that
the absence of the distinguishing Ia feature (i.e. the Si~II 6150 \AA)
during the pre maximum phase and its weakness in post maximum is
reflected in the LE spectrum. In fact, while our synthetic spectra,
calculated using ``normal'' events like SNe~1994D and 1998aq, do show a
clear imprint of the P-Cyg profile displayed by this line during the
photospheric phase, this is practically absent in the data of
1991T. Another clear difference is that visible in correspondence of
the CaII H\&K lines. Again, the weakness of this feature in 1991T is a
consequence of what happened in the pre-maximum phase, when this line
was totally absent (\citealt{filippenko}).

In the other case, the match between the synthetic and the observed LE
spectrum is fairly good but, still, the observed deviations can all be
attributed to small differences in the spectral appearance of
SN~1998bu with respect to SN~1994D (see \citealt{hernandez}).

All of this has an interesting consequence, i.e. that the maximum
phase peculiarities can be easily seen in the LE spectrum. Therefore,
not only one could classify an historical SN on the basis of its LE
spectrum alone (disentangling for example between a Type I and a Type
II), but one could also get some more detailed information about its
features during maximum light, so that it would be possible to
distinguish, for instance, between a normal Ia, a 1991T-like or a
1991bg-like event.  This might have some relevance when traces of an
historical SN explosion are found by detecting its LEs, as in the
recent cases near the SN remnant of Cas A
\citep{krause} and in LMC \citep{rest}. Provided that the LE is bright
enough to allow a spectroscopic observation, this would provide a
powerful tool to type the SN. The detections by
\citet{krause} and \cite{rest}
lets us hope that more cases will be discovered in connection with SN
remnants in our Galaxy, offering a unique chance to directly classify
the explosion that generated them and to establish a firm connection
between the two.

In principle one could envisage to use the Case A LE observations to
derive some information about the extinction law in the host
galaxy, simply comparing the SN time integrated spectrum and the LE
spectrum.  In fact, under the conservative assumption that the LE and
the SN suffer the same extinction (i.e. $\kappa(V)$=0 in
equation~(\ref{eq:law1})), the ratio between $S_{LE}$ and $\cal S$ is
reddening independent. Unfortunately, even in these simplified
circumstances, one is left with the albedo and forward scattering
degree wavelength dependence. More in general, due to the way the
effects of geometry, optical depth and dust properties combine
together, one cannot derive unambiguous information on the extinction
law since very similar solutions can be obtained using different
ingredients. Some meaningful results might be achievable from
spectroscopic observations of LEs if some of the dependencies can be
removed by other and independent observations. Of great help, in this
respect, are high spatial resolution images, which can give direct
information on the geometrical dust distribution.

With the increased interest on Type Ia SNe related to their usage in
cosmology and the availability of larger and larger observing
facilities, it is most likely that an increasing number of objects
belonging to this class will be observed in the near future and
followed up to even more advanced phases than it is currently done.
Our forecast is that these investigations will probably be disturbed
by the presence of LEs, too faint to be detected by the typical
current instrumentation.  In fact, as we have shown here, a dust
density of the order of $n_H$=5 cm$^{-3}$ is sufficient to produce a
LE compatible with those seen in SNe~1991T and 1998bu. Due to the lack
of LE detections in the vast majority of the cases \citep{boffi}, in
Paper I we had concluded that the height scale of Type Ia must be
significantly higher than that of the dust in the disk of host
galaxies. Indeed, a very similar conclusion was reached by \cite{mdv} on
the basis of completely independent considerations.

Nevertheless, when one pushes the observations at later phases, due to
the fast decay typical of these objects during the nebular phase
($\sim$5 mag yr$^{-1}$), the light curve rapidly reaches levels where
a LE produced by a much lower densities would dominate the spectrum.
For example, for the same geometry used in Figure~\ref{fig:lcsheet}, a
density of $n_H$=0.5 cm$^{-3}$ would produce a LE approximately 2.5
mag fainter, implying that the transition to a LE dominated spectrum
would have started only 6 months later.  Therefore, the study of Type
Ia at phases later than 1.5-2 yr, aimed for example at detecting
deviations from the radioactive decay, will be most probably hindered
by the emergence of a Case A LE.

If, on the one side, it is true that there is an increasing number of
instances where LEs of this kind are actually observed (see also the recent
case reported by \citealt{sugerman05} and \citealt{vandyk} for the
core-collapse SN~2003gd), one may in fact wonder whether and to which
extent early phases could be affected, modifying for instance the
light curve shape, the overall luminosity and introducing
peculiarities in the spectral appearance. As we have shown in this
paper, this can actually take place when the dust cloud is very close
to the SN and has small dimensions, circumstances that we have
generically indicated as Case B. As it has been independently shown by
\cite{wang05}, circumstellar dust can in principle alter the observed
light and colour curves producing, for example, interesting effects on
the derived extinction curve.  Our analysis (see
Secs.~\ref{sec:closedust} and
\ref{sec:caseb}), run under the simplifying hypothesis that the dust
is distributed according to a $r^{-2}$ law with an inner cavity with
radius $R_0$, shows that already with moderate optical depths
($\tau_d(V)$=1), the presence of dust in the immediate vicinity of a
Type Ia ($R_0\sim$10$^{16}$ cm) would turn into significant changes in
the photometric parameters, which would still be within the normally
observed ranges. In particular, the post maximum decline rate $\Delta
m_{15}$ would be smaller and the SN would also appear inherently
brighter, due to the additional flux contribution added by the
underlying LE, i.e. producing an additional scatter in the
Pskovski-Phillips relation. In this respect, we notice that there are
many instances of Type Ia showing low decline rates ($\Delta
m_{15}(B)<$1) without displaying the spectral peculiarities typical of
slow-declining SN~1991T-like events (see for example
\citealt{phillips}, Table~2).

Besides influencing the photometric parameters, the time integration
during the fast evolving phase around maximum light tends to produce
changes in the line widths and peaks, altering the measured expansion
velocities. As we have said, according to our simulations the
deviations from a LE-free standard behaviour would be rather subtle
and could be easily misidentified as the intrinsic features of a
peculiar object. Whilst most of the photometric parameters tend to be
mildly altered (at least for relatively low values of the dust optical
depth) the colour evolution shows quite a clear deviation from the
homogeneous behaviour shown by Type Ia between 30 and 90 days past
maximum light \citep{lira}. If indeed all LE-free objects follow that
path, then the colour curve could be a good diagnostic tool to unveil
the presence of a LE.

As pointed out by \cite{wang05}, the presence of Case B LEs could
explain the low values for $R_V$ which have been deduced from SN observations
(see for example \citealt{phillips}, \citealt{krisciunas00}, \citealt{knop},
\citealt{altavilla} and \citealt{krisciunas05} for the most recent examples). 
As we have shown here, this is in principle true, even though some
side effects should be clearly detectable, at least for the simple
dust distributions we have analyzed.

According to the lower panel of Fig.~\ref{fig:windmccol}, or to the
equivalent Fig.~3 of \cite{wang05}, one would expect that the deduced
$R_V$ remains roughly constant ($\sim$1) in this phase
range. Nevertheless, in those plots $E(B-V)$ is kept at the constant
value attained at maximum light, while the difference in colour
between the global spectrum and the SN spectrum changes with time,
both in our Monte Carlo calculations and in the semi-analytical model
presented by
\cite{wang05}. In particular, when the global
colour becomes equal to or even bluer than that of the unreddened SN,
the extinction correction is expected to produce weird results.
Therefore, if one attributes the deviations from the standard colours
to a different extinction law, she also has to accept the puzzling
conclusion that this has to change with time. We reckon that both the
colour evolution and the time variability of the derived extinction
law would be sufficient to infer that the observed deviations are not
due to a peculiar dust composition but rather to a LE.

Another interesting fact worth to be mentioned is the discrepancy that
one would find between the extinction derived from the equivalent
width of the NaI~D lines and that deduced from photometric or
spectroscopic comparisons with templates. In fact, while the intensity
of the NaI~D absorption is related only to the dust present along the
line of sight (and would hence give a direct estimate of $\tau_d$),
all other methods would be affected by the LE contribution. As we have
seen in Sec.~\ref{sec:caseb}, this tends to significantly decrease the
amount of derived extinction. In the case of normal dust-to-gas ratios
and for a simple $r^{-2}$ spherically symmetric dust geometry, this
would imply that the values deduced from the NaI doublet are
systematically higher than those inferred from other methods (compare
$A^0_V$ with $A_V$ or $A^\prime_V$ in Table~\ref{tab:models}).

In all this discussion we have assumed (and required) that there can
be dust very close to the SN. In this respect there is a very
important consideration one has to make. In fact, if on the one hand
it is true that LE effects become more subtle when the dust is closer
to the SN (and can therefore be mistaken with intrinsic properties),
on the other there actually might be a lower limit for the SN-dust
distance.  In fact, for typical expansion velocity of 10$^4$ km
s$^{-1}$, the ejecta would reach $R_0$=10$^{16}$ cm in less than four
months. This is a problem that needs to be addressed with detailed
modeling, in order to see whether the interaction would be able to
destroy and/or sweep away the dust and to evaluate possible effects on
the observed reddening, its time evolution and so on. One should also
mention that the radiation field produced by the SN explosion would
probably evaporate the dust up to a certain radius. If in the case of
a core-collapse event the UV flash is expected to produce a cavity in
the dust with a radius of $\sim$10$^{17}$ cm (see for example
\citealt{chevalier}), the lack of UV photons in a Type Ia might still
make the dust survive the explosion. Another implication that needs to
be addressed is related to the fact that substantial amounts of dust
should be associated with correspondingly high amounts of gas. If the
gas is close enough to the SN, then some interaction is expected, like
in the case of SN~2002ic
\citep{hamuy}.

Finally, what remains to be studied is the effect of asymmetric dust
distributions (disks, tori, lobes and so on) that, among other things,
would certainly produce a net continuum polarization up to several per
cent (see Paper I).  Being close to the SN the polarized flux could be
in fact high enough to produce detectable effects and serve as a
useful diagnostic tool to identify the presence of an underlying Case
B LE. Finally, to make the scenario fully self-consistent, one should
also investigate the effects of dust heating on IR re-emission, which
might turn out to be another observable side-effect produced by this
kind of LEs. 

Detailed analysis of reddened objects will certainly clarify whether
Case B LEs are indeed taking place and this, in turn, will probably
give us a few more hints about the nature of Type Ia progenitors. We
are currently working on the highly reddened SN~2003cg
\citep{nancy}, which shows clear features that could be interpreted as
the signature of a Case B LE.  The results will be reported in another
paper of this series.

\section{Conclusions}
\label{sec:conclusion}

In this paper we have analyzed and discussed the effects of LEs on light
curves, colours and spectra of Type Ia SNe for two different dust
geometries: extended clouds (Case A) and small circumstellar clouds
(Case B). The main results can be summarized as follows:

\begin{enumerate}
\item The effect of Case A LEs is limited to the late phases.
The exact time when the LE contribution dominates the global
luminosity depends on dust distance and density.
\item There is a transition phase, during which the LEs contribution
becomes more and more important and the global spectrum turns from a
normal nebular spectrum into a pure LE spectrum.
\item During the transition phase, the overall colour tends to become bluer, 
spectral features broader and emission line peaks displaced with respect to a 
pure SN nebular spectrum.
\item During the LE-dominated phase the spectrum retains the peculiarities shown by the 
SN at maximum, so that the LE detection in an historical SN allows to classify it
with a good confidence level.
\item The spectroscopic observation of Case A LEs in the vicinity of SN remnants can 
help in establishisng a firm and direct relation between the SN type and the remnant.
\item The information on the extinction properties one can derive from LEs (both Case A and B)
observations may lead to incorrect conclusions on the dust nature.
\item Case B LEs affect in principle all phases, without necessarily 
producing notable effects on the spectra. This is particularly true
if the dust is confined within $\sim$10$^{16}$ cm from the SN explosion and $\tau_d<$1.
\item In the presence of circumstellar dust, the light curves tend to become broader
with a smaller $\Delta m_{15}$. The rising time
increases, the late time decay becomes slower and the luminosity difference between maximum
light and nebular phase tends to decrease.
\item The SN absolute luminosity deduced from the apparent magnitude, colour excess and a 
standard extinction law is brighter than the real value.
\item The difference in luminosity is of the same order of the correction foreseen by the
Pskowski-Phillips relation. This, coupled to the fact that $\Delta
m_{15}$ is decreased, is a possible source of noise in the
luminosity-decline rate relation.
\item As shown by \cite{wang05}, Case B LEs modify the luminosity and colours in such a way
that the apparent selective-to-total extinction ratio $R_V$ is significantly decreased.
\item In this scenario, $R_V$ should show a rapid evolution during the first months after
explosion and this, together with a non standard colour evolution,
should be a simple diagnostic tool to unveil the underlying LE.
\end{enumerate}

%%%%%%%%%%%%%%%%%%%%%%%%%%%%%%%%%%%%%%%%%%%%%%%%%%%%%%%%%%%%%%%%%%%%%%%%%%%%

 \bsp \label{lastpage} \end{document}